# Heuristic for Edge-enabled Network Slicing Optimization using the "Power of Two Choices"

Jose Jurandir Alves Esteves*†, Amina Boubendir*, Fabrice Guillemin* and Pierre Sens†
*Orange Labs, France
†Sorbonne Université / CNRS / Inria, LIP6, France
{josejurandir.alvesesteves, amina.boubendir, fabrice.guillemin}@orange.com, pierre.sens@lip6.fr

*Abstract*—We propose an online heuristic algorithm for the problem of network slice placement optimization. The solution is adapted to support placement on large scale networks and integrates Edge-specific and URLLC constraints. We rely on an approach called the "Power of Two Choices" to build the heuristic. The evaluation results show the good performance of the heuristic that solves the problem in few seconds under a large scale scenario. The heuristic also improves the acceptance ratio of network slice placement requests when compared against a deterministic online Integer Linear Programming (ILP) solution.

## I. Introduction

By breaking the link between functions and their hosting hardware, Network Function Virtualization (NFV) deeply modifies the architecture and the operation of telecommunications networks. While so far specific network functions hardware had to be deployed, virtualized network functions (VNFs) can today thanks to virtualization be deployed on common hardware. Furthermore, the life-cycle of VNFs can be managed independently from the underlying physical infrastructure.

A network then becomes a programmable platform, which can host chains of VNFs so that various logical networks (viewed as series of VNFs) can be deployed over the same shared Physical Substrate Network (PSN). This has given rise to the concept of Network Slicing, which takes benefit of the logical and/or physical separation of network resources to allow multi-tenancy support, customization and isolation of Network Slices [1]. Even though several definitions of slicing have been considered by standardization organizations, for instance NGMN [2], 3GPP [3], or ETSI [4], we consider in this paper a network slice as a set of VNFs interconnected by transmission links. Such a Service Function Chain (SFC) has bandwidth and latency constraints in addition to traditional IT resource requirements in terms of computing and storage. All these constraints apply to Service Level Agreement (SLA) as requirements of the Network Slice tenant.

Proper management and orchestration of Network Slices, VNFs and their associated Virtual Links (VLs) are essential to meet and maintain the SLAs of concurrent Network Slice. Simultaneously achieving optimal utilization of network resources and guaranteeing SLAs of Network Slices is a major challenge for network operators. One component of this process is the placement of network slices.

Network slice placement can be viewed as an optimization problem that consists of choosing the servers of the PSN in which the VNFs composing a Network Slice can be deployed and which physical links to use in order to steer traffic between these VNFs. This problem contains a specific optimization objective (e.g., minimizing resource consumption, optimizing a specific QoS metric, etc.) that must be satisfied [5], [6]. Network slice placement problem is a special case of more general problems such VNF Forwarding Graph Embedding (VNF-FGE), Virtual Network Embedding (VNE), Service Function Chain placement (SFC-P), and VNF Placement and Chaining (VNF-PC) problems. Numerous papers about network slice placement and its variants use heuristic-based approaches to solve associated optimization problems. However, most of them do not jointly address the large scale PSNs and Network Slice Placement Requests (NSPR), neither the Edge-specific constraints and thus the direct impact on QoS metrics (notably, E2E latency). In this paper, we develop an original method of placing network slices through a heuristic based on the Power of Two Choices (P2C) algorithm [7]. We adapt the heuristic to large scale network scenarios and integrate both edge-specific constraints related to user location and strict end-to-end (E2E) latency requirements. We implement a policy for selecting servers for VNF placement that offloads edge data centers and improves Network Slice acceptance ratio in most simulation scenarios when compared to an online ILP-based placement algorithm proposed in a recent previous work [8] that we extend here. The organization of this paper is as follows. In Section II, we review related works. Section III provides assumptions and definitions of the model. Section IV introduces the problem statement and formulation. Section V introduces the proposed heuristic approach. The experiments and evaluation results are presented in Section VI. Some concluding remarks and perspectives are presented in Section VII.

## II. Related Work Analysis

In this section, we review some recent studies on the network slice placement problem. We notably consider comprehensive surveys like [26], [27], [6], [28]. Tables I and II reflect this analysis based on 3 axes: i) heuristics for network slice placement problem (Section II-A), ii) placement in large scale networks (Section II-B), and iii) Edge-specific constraints (latency and location) for slice placement (Section II-C).

TABLE I: Synthesis on heuristics for network slice placement optimization

| References | Heuristic strategies | | | Technical constraints | | | Optimization objectives | | |
|---|---|---|---|---|---|---|---|---|---|
| | Online | Dynamic | Decentralized | Resource utilization | Quality-of-Service | User location | Resource utilization | Quality-of-Service | Cost or revenue |
| [9], [10], [5], [11], [12], [13], [14] | X | | | X | | | X | | |
| [15] | | | X | X | | | X | | |
| [16] | X | X | X | X | | | X | | |
| [17] | X | X | | X | | | | | X |
| [18], [19] | | | | X | X | | X | | |
| [20], [21] | | | | X | X | | | X | |
| [22], [23], [24], [25] | X | | | X | X | X | X | | |

TABLE II: Synthesis on network slice placement optimization network scale aspects

| References | Network scale (# of nodes) | | | | |
|---|---|---|---|---|---|
| | PSN | | | NSPR | |
| | Small [4,50] | Medium [50,200] | Large [1000,5000] | Small [2,50] | Large [50,200] |
| [22], [17], [23], [25], [11], [12], [18], [15] | X | | | X | |
| [9], [16], [24], [20], [13], [19], [21], [8] | | X | | X | |
| [5] | | | X | | X |
| [14], [10] | | | X | X | |

## A. On heuristics for network slice placement optimization

*1) Heuristics Strategies:* Most of the works on network slice placement optimization propose heuristic approaches to addressing the scalability issues due to the ever-growing complexity of exact approaches like ILP. As shown in Table I, most of existing heuristics are static and use centralized strategies, mostly online algorithms, where placement requests arrive dynamically and are not known in advance as opposed to offline algorithms. We have identified only two decentralized approaches ([15] and [16]) where the decision is taken by multiple agents. In addition, we have found two dynamic approaches ([16] and [17]) where a (re)optimization of the already performed placement is possible. We propose in this paper an online, static, centralized approach.

*2) Heuristic technical constraints & optimization objectives:* As exhibited in Table I, technical constraints are related to resource utilization (available CPU and RAM capacity on a hosting node), QoS (e.g., E2E latency) and/or user location. Optimization objectives are related to resource utilization (minimization of bandwidth usage) and QoS metrics (maximization of availability). Table I also shows that most heuristics have focused on resource utilization without always integrating QoS constraints. Two kinds of approaches are used to take a QoS metric into account: setting QoS metric as a strict constraint, that is, a constraint that absolutely needs to be satisfied [21], and setting the QoS metric as an optimization target [20]. E2E latency is a performance measure to be minimized. Our approach integrate all three kind of constraints mentioned with and optimizes resource utilization.

## B. On Large scale considerations for network slice placement

Table II shows how existing heuristics support large scale for slice placement. Only small and medium scale networks seem to be mainly considered. The work [5] solves VNF-FGE in large scale network scenarios and proposes an online Eigen decomposition approach to show its scalability via simulations with PSN up to 5000 nodes. Also, [14] proposes a "boosted ILP" to solve VNF-FGE problem by sharing VNFs between different VNF-FGs and their objective is to minimize power consumption.

We evaluate our approach considering a large scale PSN and small scale NSPRs. Hence, there is clearly still a lack of heuristic approaches considering the large scale aspect for both the PSN and the NSPR.

## C. On Edge-specific constraints for network slice placement

Edge-specific constraints are usually related to latency and access delays and directly linked to user location. Most of works that consider E2E latency as a strict constraint for slice placement calculate latency between a source and a destination VNF. In an E2E Network Slicing view, however, the service users need be considered as end-points in the E2E latency calculation. Hence, taking user location into consideration is of utmost importance.

From Table I, very few works take user location into account. Most of them consider a preferred location for each VNF to be placed as an input of the problem [22], [23], [24]. This assumption has the advantage of reducing the problem complexity. However, to the best of our knowledge, the precise definition of the preferred placement location for each VNF of a Network Slice is not straightforward. Hence in our approach we do not consider this assumption.

## III. NETWORK MODEL

We introduce the components of the proposed model underlying the network slice placement problem: the PSN and Network Slice Placement Requests (NSPRs).

### A. Physical Substrate Network (PSN)

The PSN (see Figure 1) is divided into divided in three parts: the Virtualized Infrastructure (VI), the Access Network (AN) and the Transport Network (TN). The PSN is composed of the infrastructure resources needed to support the deployment of a network slice VNFs and their interconnection by Virtual Links (VLs).

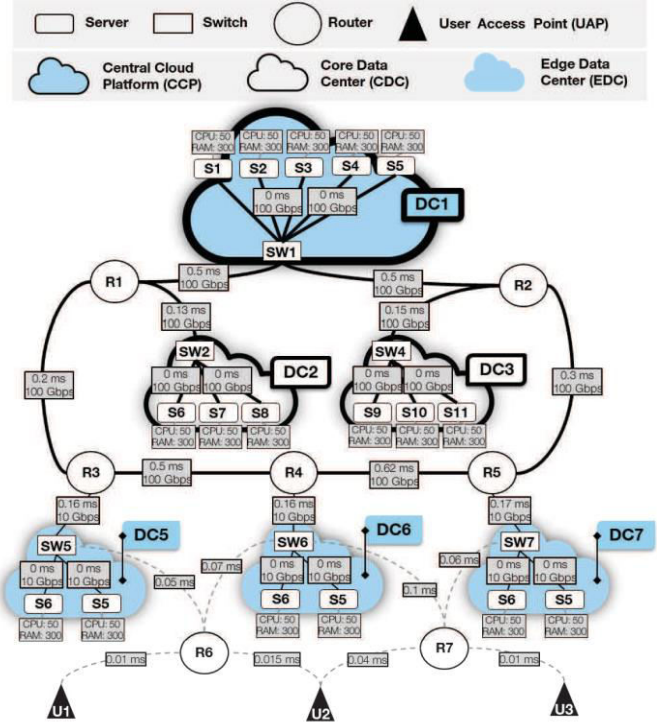

Fig. 1: Physical Substrate Network example

## B. Network Slice Placement Requests (NSPR)

A NSPR is a representation of the resource and latency requirements for a network slice to be placed on the PSN. We model an NSPR as a VNF chain (requiring IT resources) with bandwidth and latency requirements. The nodes of the NSPR are labeled with a CPU and RAM requirements (for the VNFs) and the edges are labeled with a bandwidth and latency requirement (network links).

Each group of Network Slice Users (NSUs) imposes a maximum acceptable Access latency between the UAP users are connected to and the first VNF of the NSPR they request in order to ensure the feasibility of the communication. The E2E latency requirement for each NSPR stands for the maximum latency allowed between the UAP associated to the NSPR and the last VNF of the NSPR. Figure 2 provides an example of NSPR of with VNFs and a respective UAP.

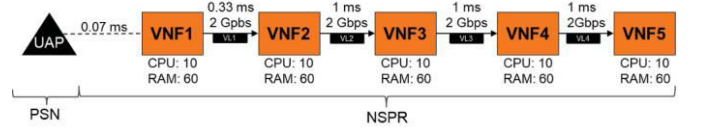

Fig. 2: NSPR example

1) *The Virtualized Infrastructure (VI):* this component of the PSN is the set of data centers (DCs) interconnected by network elements (switches and routers) and located at Point of Presence (PoP) or centralized. They offer IT resources to run VNFs.

We define three types of DCs with different capacities: Edge Data Centers (EDCs) as local DCs with small resources capacities, Core Data Centers (CDCs) as regional DCs with medium resource capacities, and Central Cloud Platforms (CCPs) as national DCs with big resource capacities.

2) *The Access Network (AN):* represents User Access Points (UAPs) (Wi-Fi APs, cellular, etc.) and Access Links. Users access the slices via one UAP, which may change during the life time of a communication by a user.

3) *The Transport Network (TN):* represents the set of routers and transmission links needed to interconnect the different DCs and the UAPs.

The complete PSN is modeled as an undirected graph, where each node has a type in the set {UAP, router, switch, server}. The nodes of type server are labeled with a CPU and RAM capacity. The edges on this graph represent the transport, access and intra-data center links. Transmission links and data center links are labeled with a bandwidth capacity and a latency. Access links are labeled with a latency. The latency between UAP and DC is referred to as Access latency. We assume the latency between two VNFs of the same DC is negligible, i.e., latency in data center links are set equal to 0.

## IV. PROBLEM STATEMENT & FORMULATION

We present here the online network slice placement problem statement and its formulation as an ILP.

### A. Online Network Slice Placement: Problem Statement

The online network slice placement problem is as follows with notation given in Tables III and IV for PSN and NPSR, respectively.

- *Given:* a NSPR to be placed and a PSN,
- *Find:* in which server $s \in S$ of the PSN to instantiate each requested VNF $v \in V$; which physical links $(a, b) \in L$ to use in order to realize the VLs $(\overline{a}, \overline{b}) \in E$ requested between these VNFs,
- *Subject to:* the servers CPU available capacity $cap_s^{cpu}, \forall s \in S$, the servers RAM available capacity $cap_s^{ram}, \forall s \in S$, the physical links bandwidth available capacity $cap_{(a,b)}^{bw}, \forall (a,b) \in L$, the access latency requirements $d^a$ requirement, and the E2E latency requirement $\delta$
- *Objective:* minimizing total resource utilization or minimizing blocking ratio.

### B. Problem Formulation

To formulate the optimization problem, we introduce the decision variables and we identify the constraints, which has to be satisfied by the placement algorithm.

TABLE III: PSN parameters

| Parameter | Description |
|---|---|
| $N$ | Network nodes (switch, servers, routers) |
| $S \subset N$ | Set of servers |
| $DC$ | Set of data centers |
| $S_{dc} \subset S, \forall dc \in DC$ | Set of servers in data center $dc$ |
| $SW_{dc}, \forall dc \in DC$ | Switch of data center $dc$ |
| $L = \{(a,b) \in N \times N \wedge a \neq b\}$ | Set of physical links |
| $cap^{bw}_{(a,b)} \in \mathbb{R}, \forall (a,b) \in L$ | Bandwidth capacity of physical link $(a,b)$ |
| $cap^{cpu}_s \in \mathbb{R}, \forall s \in S$ | CPU capacity of server $s$ |
| $cap^{ram}_s \in \mathbb{R}, \forall s \in S$ | RAM capacity of server $s$ |
| $\delta_{(a,b)} \in \mathbb{R}, \forall (a,b) \in L$ | latency induced by physical link $(a,b)$ |

TABLE IV: NSPR parameters

| Parameter | Description |
|---|---|
| $V$ | Set of VNFs of the NSPR |
| $E = \{(\bar{a},\bar{b}) \in N \times N \wedge \bar{a} \neq \bar{b}\}$ | Set of VLs of the NSPR |
| $v_{root} \in V$ | Root VNF of the NSPR |
| $d^{cpu}_v \in \mathbb{R}$ | CPU requirement of VNF $v$ |
| $d^{ram}_v \in \mathbb{R}$ | RAM requirement of VNF $v$ |
| $d^{bw}_{(\bar{a},\bar{b})} \in \mathbb{R}$ | Bandwidth requirement of VL $(\bar{a},\bar{b})$ |
| $d^{\delta}_{(\bar{a},\bar{b})} \in \mathbb{R}$ | Latency requirement of VL $(\bar{a},\bar{b})$ |
| $\delta \in \mathbb{R}$ | E2E latency requirement of the NSPR |
| $a_{max} \in \mathbb{R}$ | Access latency requirement of the NSPR |
| $a_{dc} \in \mathbb{R}$ | Access latency to data center $dc$ |

*1) Decision Variables:* We use the two following binary decision variables:

- $x^v_s \in \{0,1\}$ for $v \in V$ and $s \in S$ is equal to 1 if the VNF $v$ is placed onto server $s$ and 0 otherwise
- $y^{(\bar{a},\bar{b})}_{(a,b)} \in \{0,1\}$ for $(\bar{a},\bar{b}) \in E$ and $(a,b) \in L$ is equal to 1 if the virtual link $(\bar{a},\bar{b})$ is mapped onto physical link $(a,b)$ and 0 otherwise

*2) Problem Constraints:*

*a) VNF placement:* The following constraint ensures that 1) all VNFs of the NSPR must be placed and 2) each VNF must be placed in only one server:

$$\forall v \in V, \sum_{s \in S} x^v_s = 1 \quad (1)$$

*b) Network Resource Capacities Constraints:* Constraints (2) and (3) below ensure that the resource capacities of each server (for CPU and RAM, respectively) are not exceeded; the subsequent constraint (4) guarantees that the bandwidth capacity of each physical link is not exceeded:

$$\forall s \in S, \sum_{v \in V} d^{cpu}_v x^v_s \leq cap^{cpu}_s \quad (2)$$

$$\forall s \in S, \sum_{v \in V} d^{ram}_v x^v_s \leq cap^{ram}_s \quad (3)$$

$$\forall (a,b) \in L, \sum_{(\bar{a},\bar{b}) \in E} d^{bw}_{(\bar{a},\bar{b})} y^{(\bar{a},\bar{b})}_{(a,b)} \leq cap^{bw}_{(a,b)} \quad (4)$$

*c) Eligible Physical Path Calculation:* Constraints (5), (6) and (7) below ensure that if two connected VNFs are mapped onto different servers, the VL that connects them is mapped onto one physical path between these two servers: for all $a \in S$ and $(\bar{a},\bar{b}) \in E$,

$$\sum_{\substack{b \in N:\\(a,b)\in L}} y^{(\bar{a},\bar{b})}_{(a,b)} - \sum_{\substack{b \in N:\\(b,a)\in L}} y^{(\bar{a},\bar{b})}_{(b,a)} = x^{\bar{b}}_a - x^{\bar{a}}_a, \quad (5)$$

and

$$\forall a \in N \setminus S, \forall (\bar{a},\bar{b}) \in E, \sum_{\substack{b \in N:\\(a,b)\in L}} y^{(\bar{a},\bar{b})}_{(a,b)} - \sum_{\substack{b \in N\\(b,a)\in L}} y^{(\bar{a},\bar{b})}_{(b,a)} = 0, \quad (6)$$

$$\forall (\bar{a},\bar{b}) \in E, \forall (a,b) \in L, y^{(\bar{a},\bar{b})}_{(a,b)} + y^{(\bar{a},\bar{b})}_{(b,a)} \leq 1. \quad (7)$$

*d) Network Slice Latency Requirements Constraints:* Constraint (8) below guarantees that the latency requirements of each virtual link of the NSPR will be respected. Constraint (9) ensures that the access latency is respected and Constraint (10) reflects the fulfillment of the required E2E latency:

$$\forall (\bar{a},\bar{b}) \in E, \sum_{(a,b) \in L} \delta_{(a,b)} y^{(\bar{a},\bar{b})}_{(a,b)} \leq d^{\delta}_{(\bar{a},\bar{b})}, \quad (8)$$

$$\sum_{s \in S} a^r_s x^{v_{root}}_s \leq a_{max}, \quad (9)$$

$$\sum_{s \in S} a_s x^{v_{root}}_s + \sum_{(a,b) \in L} \sum_{(\bar{a},\bar{b}) \in c} \delta_{(a,b)} y^{(\bar{a},\bar{b})}_{(a,b)} \leq \delta. \quad (10)$$

*3) Objective Function:* We consider two objective functions. The first one is to minimize the consumption of resources. The second one is the maximization of accepted requests.

*a) Minimization of the total resource utilization:* The placement of all VNFs of a slice is mandatory otherwise the solution would violate Constraint (1). The optimization objective in this case is the minimization of bandwidth resources utilization given by

$$\min_{x,y} \sum_{(\bar{a},\bar{b}) \in E} \sum_{(a,b) \in L} y^{(\bar{a},\bar{b})}_{(a,b)} cap^{bw}_{(a,b)} \quad (11)$$

*b) Maximization of accepted slice requests:* The maximization of accepted slices requests objective function is given by Equation (12). Auxiliary variable $z \in \{0,1\}$ representing whether the NSPR is accepted ($z = 1$) or not ($z = 0$) is used in this case.

$$\max_{x,y,z} z \quad (12)$$

where the additional Constraints (13) and (14) below need to be inserted in the model:

$$\forall v \in V, z \leq \sum_{s \in S} x^v_s, \quad (13)$$

$$z \geq \sum_{s \in S} \sum_{v \in V} x^v_s - |V - 1|. \quad (14)$$

It turns out that the ILP optimization problems formulated above are very time and resource consuming to solve. This is why we introduce a heuristic approach in the following section.

## V. PROPOSED HEURISTIC ALGORITHM

We here describe the proposed heuristic based on P2C to solve the network slice placement problem introduced in Section IV. P2C has shown good results in solving a similar problem in a previous work [7] what motivates its use here.

### A. Proposed Network Slice Placement Optimization Heuristic

Algorithm 1 presents the pseudo code for the proposed heuristic to solve network slice placement problem introduced in the previous section. The algorithm is fed with the *NSPR* to place, the *PSN* and also a *policy id* parameter used to differentiate two possible candidate server selection policies.

It returns the *status* of the *NSPR* (Accepted or Rejected) and, if *status = Accepted*, it also returns the amount of bandwidth consumed $C$ by the *NSPR* and the values for $x$ and $y$ decision variables. The algorithm performs a sequence of steps for each VNF $v$ of the *NSPR*. Step 1 (line 5) calculates the set $S'$ of feasible servers for the placement of VNF $v$. This is done by the procedure getFeasibleServers detailed in Section V-B. If there are feasible servers for the placement of VNF $v$, i.e., $S' \neq \emptyset$, the algorithm proceeds to Step 2 (line 7) that returns two candidate servers $s_1$ and $s_2$ for the placement of VNF $v$ using the getTwoCandidateServers procedure detailed in Section V-C. Finally, the algorithm proceeds to the Step 3 (lines 8-53) evaluate candidate servers $s_1$ and $s_2$. If one of these two servers was previously used to place VNF $v-1$, i.e., it is the same than *last_s*, this server is chosen for the placement of VNF $v$ as it is the optimal solution with 0 bandwidth consumption. Otherwise one path $P_i$ needs to be calculated between *last_s* and $s_i$, for $i = 1, 2$. This is done using the dijkstra procedure (line 17). This latter procedure implements the Dijkstra shortest path algorithm and first tries to calculate the feasible path providing minimum bandwidth consumption between *last_s* and the evaluated server $s_i$, $i = 1, 2$. If the path satisfies the latency constraint between VNFs $v-1$ and $v$, it is returned by the dijkstra procedure. Otherwise the procedure tries to find a feasible path minimizing the latency between VNFs $v-1$ and $v$. If no feasible path is found the procedure returns $\emptyset$. The server allowing minimal embedding cost, i.e., the one which induces the use of less bandwidth resources, is chosen for the placement of VNF $v$. Variables $x$ and $y$, the available server and physical links capacities and the *NSPR* placement cost $C$ are updated accordingly. If for some VNF $v$ no feasible servers or paths are found, there is a blocking and the algorithm returns *Rejected* status (lines 52 and 56).

### B. Calculation of Eligible Servers for Placement

The procedure getFeasibleServers implements the different problem constraints described in Section IV to filter eligible servers $S'$ for the placement of VNF $v$. The implementation of these constraints is different according to the value of $v$. The value $v = 1$ corresponds to the root VNF of the *NSPR*. The procedure first obtains the eligible data centers $DC'$, i.e., the ones satisfying the access latency requirement of the *NSPR*. Two conditions are used to define whether a server

---

**Algorithm 1:** Heuristic for Network Slice Placement Optimization using Power of two Choices (P2C).

Data: $NSPR$, $PSN$, $policy\_id$
Result: $status$, $C$, $x$, $y$
1  $last\_s \leftarrow 0$, $C \leftarrow 0$;
2  $x_{v,s}^v = 0$, $\forall v \in V$, $\forall s \in S$;
3  $y_{(\bar{a},\bar{b})}^{(a,b)} = 0$, $\forall (\bar{a},\bar{b}) \in E$, $\forall (a,b) \in L$;
4  for $v \in V$ do
5    $S' \leftarrow$ getFeasibleServers($NSPR, PSN, v$) // Step 1
6    if $S' \neq \emptyset$ then
7      $s_1, s_2 =$ getTwoCandidateServers($S', v, policy\_id$) // Step 2
8      if $v = 1$ or $last\_s = s_1$ then
9        $x_{v,s_1} = 1$;
10       $last\_s = s_1$;
11       $cap_{s_1}^j -= d_v^j$, $\forall j \in \{cpu, ram\}$;
12     else if $last\_s = s_2$ then
13       $x_{v,s_2} = 1$, $\forall s \in S' \setminus s_2$;
14       $last\_s = s_2$;
15       $cap_{s_2}^j -= d_v^j$, $\forall j \in \{cpu, ram\}$;
16     else
17       $P_i =$ dijkstra($last\_s, s_i$), $i = 1, 2$;
18       if $P_1 \neq \emptyset$ and $P_2 \neq \emptyset$ then
19         $cost_i = |P_i| d_{v-1,v}^{bw}$, $i = 1, 2$;
20         if $cost_1 \leq cost_2$ then
21           $x_{v,s_1} = 1$;
22           $last\_s = s_1$;
23           $cap_{s_1}^j -= d_v^j$, $\forall j \in \{cpu, ram\}$;
24           $y_{(a,b)}^{(v-1,v)} = 1$, $\forall (a,b) \in P_1$;
25           $cap_{(a,b)}^{bw} -= d_{(v-1,v)}^{bw}$, $\forall (a,b) \in P_1$;
26           $C += cost_1$;
27         else
28           $x_{v,s_2} = 1$;
29           $last\_s = s_2$;
30           $cap_{s_2}^j -= d_v^j$, $\forall j \in \{cpu, ram\}$;
31           $y_{(a,b)}^{(v-1,v)} = 1$, $\forall (a,b) \in P_2$;
32           $cap_{(a,b)}^{bw} -= d_{(v-1,v)}^{bw}$, $\forall (a,b) \in P_2$;
33           $C += cost_2$;
34       else if $P_1 \neq \emptyset$ then
35         $x_{v,s_1} = 1$;
36         $last\_s = s_1$;
37         $cap_{s_1}^j -= d_v^j$, $\forall j \in \{cpu, ram\}$;
38         $cost_1 = |P_1| d_{v-1,v}^{bw}$;
39         $y_{(a,b)}^{(v-1,v)} = 1$, $\forall (a,b) \in P_1$;
40         $cap_{(a,b)}^{bw} -= d_{(v-1,v)}^{bw}$, $\forall (a,b) \in P_1$;
41         $C += cost_1$;
42       else if $P_2 \neq \emptyset$ then
43         $x_{v,s_2} = 1$;
44         $last\_s = s_2$;
45         $cap_{s_2}^j -= d_v^j$, $\forall j \in \{cpu, ram\}$;
46         $cost_2 = |P_2| d_{v-1,v}^{bw}$;
47         $y_{(a,b)}^{(v-1,v)} = 1$, $\forall (a,b) \in P_2$;
48         $cap_{(a,b)}^{bw} -= d_{(v-1,v)}^{bw}$, $\forall (a,b) \in P_2$;
49         $C += cost_2$;
50       else
51         • Backtrack to initially available PSN capacities;
52         $x \leftarrow \emptyset$; $y \leftarrow \emptyset$; $status = Rejected$;
53         return;
54   else
55     • Backtrack to initially available PSN capacities;
56     $x \leftarrow \emptyset$; $y \leftarrow \emptyset$; $status = Rejected$; return;
57 $status = Accepted$;

---

$s$ located in a data center $dc \in DC$ is eligible for the placement of VNF 1:

1) Server $s$ has enough CPU and RAM resources to host VNFs 1 and 2. In this case, the available bandwidth capacity of the physical link connected to this server does not need to be checked since the server can host both VNFs 1 and 2 without using any bandwidth.
2) Server $s$ has enough CPU and RAM resources available to

host VNF 1 only. In this case, if server $s$ were selected to place VNF 1, VNF 2 would need to be placed on a server $s' \neq s$. Hence, to be considered eligible, the data center link connected to server $s$ must have enough available bandwidth capacity to host VL (1,2).

Assume now that $v = |V|$, where $|V|$ is the length of the NSPR in number of VNFs to be placed. This means that the last VNF of the *NSPR* is to be placed. The procedure iterates through all servers in the network to find the feasible ones. The server *last s* is eligible if it has enough CPU and RAM capacities to host VNF $|V|$. Other servers $s$ located in *last dc* will be eligible if they have enough CPU and RAM capacities to host VNF $|V|$ and if there is an intra-data center path between *last s* and $s$ with enough available bandwidth capacity to steer traffic between VNFs $|V|-1$ and $|V|$. A server $s$ in a data center $dc \neq last\ dc$ is considered eligible if there is feasible path between *last s* and $s$ to map VL $(|V|-1, |V|)$, i.e., a path respecting the latency requirements and bandwidth requirements of the VL $(|V|-1, |V|)$.

If $1 < v < |V|$, the procedure also iterates through all servers in the network to find the eligible ones but the conditions to determine if a server is eligible are different from the case when $v = |V|$. Two conditions are used to define if server *last s* is eligible: 1) *last s* has enough CPU and RAM capacity to host VNF $v$ and VNF $v+1$. In this case the available bandwidth capacity of the physical link connected to *last s* does not need to be checked since the server can host both VNFs without using any bandwidth; 2) *last s* has only enough CPU and RAM capacity to host VNF $v$ and there is an intra-data center path between *last s* and $s$ with enough available bandwidth capacity to steer traffic between VNFs $v$ and $v+1$. In this case, if server *last s* were selected to place VNF $v$, VNF $v+1$ would need to be placed in a server $s' \neq last\ s$ hence the data center link connected to server *last s* must have enough available bandwidth capacity to the VL between VNFs $v$ and $v+1$.

Two conditions are used to define if a server $s \neq last\ s$ located in *last dc* is eligible: $s$ has enough CPU and RAM capacities to host VNF $v$ and VNF $v+1$ and there is an intra-data center path between *last s* and $s$ with available bandwidth capacity to serve VL $(v-1, v)$; $s$ has enough CPU and RAM capacities to host only VNF $v$. In the last case there must be an intra-data center path between *last s* and $s$ with enough available bandwidth capacity to serve VL $(v-1, v)$ and the data center link connected $s$ must have enough available bandwidth capacity to serve VL $(v, v+1)$. For a server $s$ in a data center $dc \neq last\ dc$ to be eligible, it must exist a feasible
path between *last_s* and $s$ to map VL $(v-1, v)$, i.e., there is a path that respects the latency requirements and bandwidth requirements of the VL $(v-1, v)$.

*C. Selection Policies of Candidate Placement Servers*

The `getTwoCandidateServers` procedure receives as arguments the set $S'$ and the *policy id* parameter and returns two candidate servers $s_1$ and $s_2$ selected using the server selection policy coded by *policy id* parameter.

*1) Policy 1:* This server selection policy chooses $s_1$ and $s_2$ completely randomly from $S'$. If $|S'| = 1$, we set $s_2 = s_1$.

*2) Policy 2:* This is a more intelligent policy. It also selects $s_1$ and $s_2$ randomly but preferentially from CCPs or CDCs. The procedure will try to select $s_1$ and $s_2$ from a subset of $S'$ containing only servers located in CCPs. If it is not possible it will try servers located in the CDCs and in the last case it will try to select EDCs servers. The aim of this policy is to save EDCs capacities when possible since these are critical resources.

VI. IMPLEMENTATION & EVALUATION RESULTS

We present in this section the implementation and the numerical experiments carried out to evaluate the heuristic.

*A. Implementation Details and Experimentation Settings*

We have implemented the proposed heuristic and ILPs in Julia. We used the default branch-and-bound algorithm from `ILOG` CPLEX 12.9 solver to solve the ILPs. Experiments were executed in a 2x6 cores @2.95Ghz 96GB machine.

*1) Physical Substrate Network Settings:* We considered a PSN that could reflect that of an operator such as Orange, see [29]. In this network, 3 types of DCs match our description made in Section III. Each CDC is connected to 3 EDCs which are 100km away. CDCs are interconnected and connected to a CCP that is 300 km away. Tables V and VI summarize the DCs and transport links properties. The CPU and RAM capacities of each server are 50 and 300 units, respectively. Latency is computed by considering the speed of light in fiber.

TABLE V: Data centers description

| Data center type | Number of data centers | Number of servers per data center | Intra data center links BW capacity |
|---|---|---|---|
| CCP | 1 | 16 | 100 Gbps |
| CDC | 5 | 10 | 100 Gbps |
| EDC | 15 | 4 | 10 Gbps |

TABLE VI: Transport links capacities

| | CCP | CDC | EDC |
|---|---|---|---|
| CCP | NA | 100 Gbps | 100 Gbps |
| CDC | 100 Gbps | 100 Gbps | 100 Gbps |
| EDC | 10 Gbps | 10 Gbps | 10 Gbps |

*2) Network Slice Placement Requests Settings :* Tables VII and VIII show the network resources and latency requirements for the three NSPR classes taken into account: Best Effort (BEF), Ultra Reliable Low Latency Communications (URLLC) and Enhanced Mobile Broadband (eMBB).

*3) Tested Algorithms:* We compare four algorithms: two variants of the ILP introduced in Section IV (ILP 1 and 2 for objective functions $a$ and $b$, respectively) and two variants of the proposed heuristic approach introduced in Section V (P2C 1 and P2C 2 for server selection policies 1 and 2, respectively).

*4) Simulation Scenarios:* We consider three simulation scenarios named BEF, URLLC, eMBB in which all requests to be placed are of the same class and one simulation scenario named MIX in which we have a percentage of requests of each class: 67% of BEF, 22 % of eMBB and 11% of URLLC. We set simulation duration to 2000 time units.

TABLE VII: Resource requirements by NSPR class

| Request class | CPU requested by each VNF | RAM requested by each VNF | BW requested by each VL |
|---|---|---|---|
| Best Effort | 10 | 60 | 1 |
| uRLLC | 15 | 90 | 1 |
| eMBB | 25 | 150 | 2 |

TABLE VIII: Latency requirements by NSPR class

| Request class | Access delay requirement | Delay requirement for VL1 | Delay requirement for VL2 | Delay requirement for VL3 | Delay requirement for VL4 |
|---|---|---|---|---|---|
| uRRLC | 0.03ms | 0.33ms | 0.33ms | 0.33ms | 0.33ms |
| eMBB | 0.07ms | 0.33ms | 1ms | 1ms | 1ms |
| Best Effort | 0.07ms | 0.67ms | 1ms | 1.33ms | 1.33ms |

### B. Network Load Calculation

We use the formula proposed in [29] to calculate the arrival rates of NSPRs ($\lambda^k$) in the different network load conditions used: underload ($\rho < 1$), critical ($\rho = 1$), and overload ($\rho > 1$). Network loads are calculated using CPU resource since it is the scarcest resource in the PSN. Generally speaking, we set $1/\mu^k = 100$ time units for all $k \in K$, the set of slice classes. For resource $j$ with total capacity $C_j$, the load is $\rho_j = \frac{1}{C_j} \sum_{k=1}^{K} \frac{\lambda^k}{\mu^k} A_j^k$, where $A_j^k$ is the number of resource units requested by a NSPR of class $k$.

### C. Evaluation Metrics

We consider 3 performance metrics:
1) Average execution time: the average execution time in seconds needed to place 1 NSPR;
2) Average final blocking ratio: the average of the final blocking ratios. The final blocking rations are calculated as $\frac{\text{\# accepted NSPRs}}{\text{\# of NSPR's arrivals}}$ at the end of each execution. We average the results for 100 executions;
3) Resource utilization: the amount CPU, RAM and bandwidth consumed.

### D. Average Execution Time Evaluation

The average execution times in function of the number of servers in the PSN is given in Figure 3. Starting from a PSN with 126 servers as described in Section VI-A1 and captured in Figure 4 we generated new PSN settings by doubling the number of servers in each DC. The evaluation results confirmed our expectations showing that the average execution time grows much faster for the ILPs than for the heuristics. In the scenario with 16128 nodes the execution times are 9.8 and 12.5 seconds for the ILPs 1 and 2 respectively and 2.17 and 1.96 seconds for P2C 1 and 2 respectively.

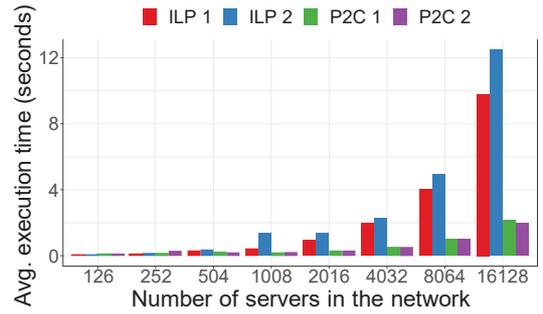

Fig. 3: Average execution time evaluation.

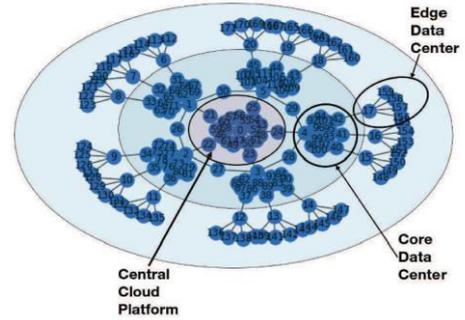

Fig. 4: Used Physical Substrate Network topology

### E. Average Final Blocking Ratio Evaluation

Figure 5 shows the blocking ratios results. We expected ILP 2 would provide the best acceptance ratio results which was not confirmed: ILP 2 obtains the best acceptance ratio in scenarios BEF, eMBB, and MIX, but in URLLC scenario, P2C 1 and 2 provide the best acceptance ratio. We explain this by a critical load balancing as heuristics usually perform better load balancing (see Figures 8-11). Also, ILP 1 has good performance in the eMBB scenario as shown in Figure 5(b) as the bandwidth here is a critical resource so ILP 1 resolve with optimal bandwidth consumption by concentrating the VNFs on the same machines. However ILP 1 have higher final blocking ratio than P2C in BEF, MIX and URLLC simulation scenarios (see Figures 5(a), 5(c) and 5(d)) since the strategy of concentrating VNFs in the same machines does not fit well. In the BEF simulation scenario, since we have only BEF NSPRs which have low CPU and RAM requirements, the ILP 1 strategy ends up concentrating all the VNFs of each NSPR in the same machine. Since the first VNF of the NSPR is always placed in a EDC due to the access latency requirements, the optimal solution considering bandwidth minimization is to concentrate all the VNFs in the same machines on a EDC. This strategy leads to a overload of EDC servers and without EDC servers available the PSN cannot accept new NSPRs. In URLLC and MIX simulation scenarios, the same problem happens but with lower intensity since the CPU and RAM required by the VNFs in these cases are higher than in the BEF simulation scenario which reduces the concentration of VNFs in the same machine.

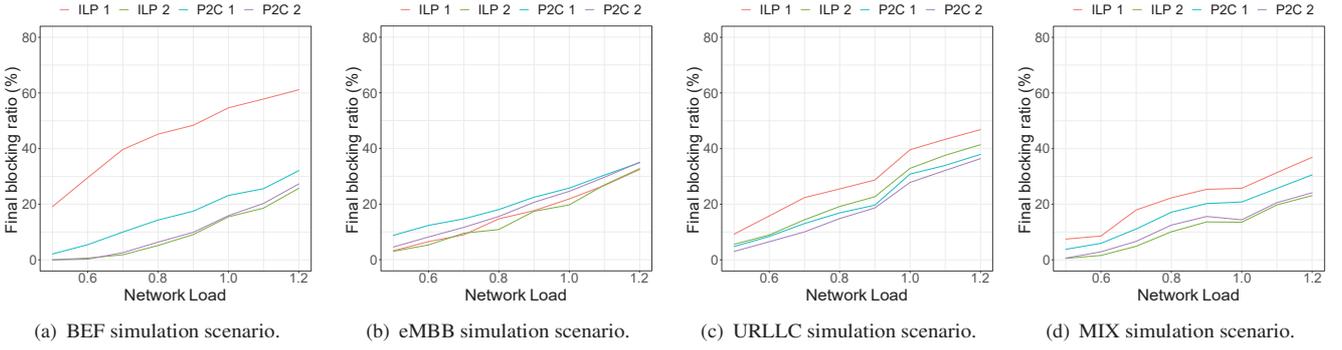

(a) BEF simulation scenario.  (b) eMBB simulation scenario.  (c) URLLC simulation scenario.  (d) MIX simulation scenario.

Fig. 5: Average final blocking ratios.

The random selection policies of candidate placement servers implemented in P2C 1 and 2 helps improve load balancing and avoids EDC overload. P2C 2 has the best performance when compared with P2C 1 as it seeks to offload EDC critical resources by placing VNFs in CCP or CDC servers preferentially. Figures 6 and 7 present how much each VNF participates in the blocking ratio obtained with P2C 1 and P2C 2 respectively in the URLLC scenario. We see that P2C 2 highly reduces the amount of blocking in the root VNF of the NSPRs and provides a well balanced blocking profile.

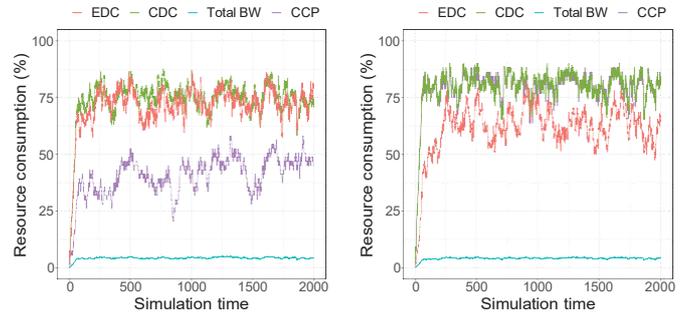

Fig. 8: P2C 1 Resource cons.  Fig. 9: P2C 2 Resource cons.

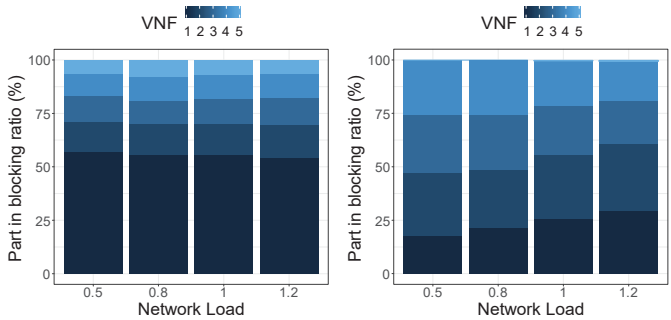

Fig. 6: Blocking ratio: P2C 1.  Fig. 7: Blocking ratio: P2C 2.

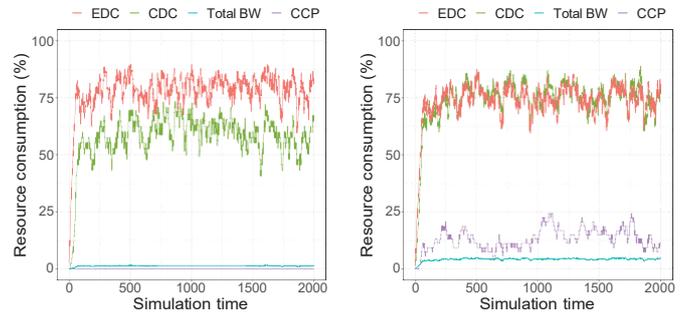

Fig. 10: ILP1 Resource cons.  Fig. 11: ILP2 Resource cons.

*F. Resource Utilization Evaluation*

Figures 8-10 show the amount of resources consumed during one simulation for P2C 1, P2C 2 and ILP algorithms, respectively. In this simulation, we consider the URLLC simulation scenario with a critical network load ($\rho = 1$). As expected, the graphics show that the P2C heuristics distributes better the load among data centers and consumes more bandwidth resources. P2C 1 distributes almost equally the load between EDCs, CDCs and the CCP, while P2C 2 concentrates the load on CCP and CDCs to offload EDCs. The ILP 1 concentrates the load on EDCs and CDCs to minimize bandwidth consumption, while the ILP 2 also uses uses CCP resources. The total bandwidth consumption obtained with the ILP 1 is minimum since it calculates solutions with optimal bandwidth consumption. ILP 2 objective function leads to solution higher bandwidth consumption than ILP 1.

## VII. CONCLUSION & PERSPECTIVES

We have proposed an online heuristic to optimize Network Slice Placement with three main contributions: i) adaption to slice placement requests on large scale networks, ii) integration of Edge-specific and URLLC-based QoS constraints (E2E latency), iii) reuse of the strength of P2C algorithm to implement selection policies. Evaluation results show that the heuristic yields good solutions within a small execution time (1.96s for a PSN of 16128 nodes). The selection policies improve load balancing and reduce load of edge data centers which improves the acceptance ratio in most simulation scenarios comparing to an online ILP-based placement algorithm. As a future work, we plan to extend this approach to support placement over multiple network domains and explore the use of machine learning strategies for network slice placement optimization.

ACKNOWLEDGMENT

This work is in the framework of 5GPPP MON-B5G project (www.monb5g.eu). The experiments were conducted using Grid'5000 a large scale testbed by INRIA and Sorbonne University (www.grid5000.fr).